\documentclass{PoS}

\title{Study of $a_{0}(980)-f_{0}(980)$ mixing at BES~III and study of charged $\kappa$ at BES~II}

\ShortTitle{Study of light scalar mesons at BES}

\author{\speaker{Beijiang Liu~(for BES~III collaboration)}\\
        The Chinese University of Hong Kong,
Shatin N.T., Hong Kong.\\
The University of Hong Kong, Pokfulam Road, Hong Kong.\\
        E-mail: \email{liubj@mail.ihep.ac.cn}}

\abstract{Recent BES results on light scalars are reported in this
talk, including the observation of a charged $\kappa^{\pm}$ decaying
to $K^{\pm}\pi^0$ with $5.8 \times 10^7 $ $J/\psi$ data at BES~II
and the direct measurements of $a^{0}_{0}(980)-f_{0}(980)$ mixing in
the processes $J/\psi\to\phi f_{0}(980)\to\phi
a^{0}_{0}(980)\to\phi\eta\pi^{0}$ and $\chi_{c1}\to\pi^{0}
a^{0}_{0}(980)\to\pi^{0} f_{0}(980)\to\pi^{0}\pi^{+}\pi^{-}$ with
$2.26 \times 10^8$ $J/\psi$ data and $1.06 \times 10^8$
$\psi^{\prime}$ data at BES~III.}

\FullConference{35th International Conference of High Energy Physics - ICHEP2010,\\
        July 22-28, 2010\\
        Paris France}

\begin{document}

\section{Introduction}
There has been much argument whether $\sigma$  and $\kappa$  exist,
due to the facts that the total phase shifts in the lower mass
region are much less than 180 degrees and they do not fit into
ordinary meson nonets. For $f_0(980)$ and $a_0(980)$, whether they
are $q\bar{q}$ mesons, 4-quark states , hybrids or $K \bar{K}$
molecules is also controversial~\cite{nonqq,lhs}. The study of their
nature has been one of the important topics in the light hadron
spectroscopy. BEPC~II/BES~III~\cite{bes3} is a major upgrade of the
BESII experiment at the BEPC accelerator~\cite{bes2} for studies of
hadron spectroscopy and $\tau$-charm physics \cite{bes3phys}. In
this talk, we present recent results from the study of these light
scalars at BES~II and BES~III.

\section{Study of charged $\kappa$ at BES~II}
The $\sigma$ and $\kappa$ were first found in the analysis of
$\pi\pi$ and $\pi K$ scattering data, and they can not be filled
into any nonets of ordinary $q\bar q$ meson. Evidences for the
neutral $\kappa$ have been reported by E791\cite{Aitala:2002kr} and
FOCUS\cite{Link:2002ev} experiment. In 2006, BESII reported the
neutral $\kappa$ in the decay of
$J/\psi\to\bar{K^\star}(892)^0K^+\pi^-$\cite{Ablikim:2005ni}. The
existence of a neutral $\kappa$ motivates the search for a charged
partner. In this proceeding, we present the search for a charged
$\kappa$ in $J/\psi\to K^{\pm}K_S\pi^{\mp}\pi^0$ at
BES~II~\cite{Ablikim:2010kd}.

\begin{figure}[htbp]
       \centerline{\includegraphics[height=4 cm,width=4 cm]
                                   {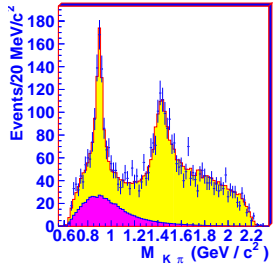}}
   \caption{Invariant mass $m(K^+\pi^0) + c.c.$
from $J/\psi\to K^*{(892)}^+ \kappa^-\to K_S^0 \pi^+ K^- \pi^0$
decays reconstructed by BES~II. The crosses represent data and the
histogram is the fit result from a partial wave analysis.}
   \label{kappakpi}
   \end{figure}

Fig.~\ref{kappakpi} shows the projected invariant mass of
$K^*{}^{\pm}\pi^0$ of the selected $J/\psi\to K^*{(892)}^+
\kappa^-\to K_S^0 \pi^+ K^- \pi^0$ events. Besides strong
contributions from $K^*{(892)}^{\pm}$, $K^*{(1410)}^{\pm}$ and
$K^*{(1430)}^{\pm}$, a significant $J^P=0^+$ low mass component is
needed to describe the data. The partial wave analysis yields the
pole position of that $0^+$ component displayed in dark color
$m-i\frac{\Gamma}{2}=(849\pm 1^{+14} _{-28})-i(288\pm 01^{+64}
_{-30})$ ~MeV/c$^2$.

This result is in agreement with a recent CLEO analysis of the
resonance structure in $D^0\to K^+K^-\pi^0$ decays
\cite{Cawlfield:2006hm}, which suggests a $\kappa^{\pm}$ component
with parameters $m = (855 \pm 15)$~MeV/$c^2$ and $\Gamma = (251\pm
48)$~MeV/$c^2$. Moreover the results are in reasonable agreement
with the properties of the neutral $\kappa^0$.

\section{Study of $a_0(980)$-$f_0(980)$ mixing at BES~III }
The mixing between $a^{0}_{0}(980)$ and $f_{0}(980)$ is expected to
shed light on the nature of these two
resonances~\cite{NN_PRB88,PRD_76_074028,NN_PLB534,BK_PRC62,NN_PRL92,FE_PLB489,
wujj1,wujj2}. The $a^{0}_{0}(980)-f_{0}(980)$ mixing intensity has
been predicted to be with a larger uncertainty by various
theoretical models. No firm experimental result was
available~\cite{wujj1,wujj2}. The leading contribution to the
isospin-violating mixing transition amplitudes for $f_{0}(980)\to
a^{0}_{0}(980)$ and $a^{0}_{0}(980)\to f_{0}(980)$ is shown to be
dominated by the difference of the unitarity cut which arises from
the mass difference between the charged and neutral kaons. As a
consequence, a narrow peak of about 8~MeV$/c^2$ is predicted between
the charged and neutral kaon
thresholds~\cite{PRD_76_074028,wujj1,wujj2}. Using the samples of
$226$ million $J/\psi$ events and $106$ million $\psi^{\prime}$
events collected with the BES~III detector in 2009, we perform
direct measurements of $a^{0}_{0}(980)- f_{0}(980)$ mixing via the
processes $J/\psi\to\phi f_{0}(980)\to\phi
a^{0}_{0}(980)\to\phi\eta\pi^{0}$ and $\chi_{c1}\to\pi^{0}
a^{0}_{0}(980)\to\pi^{0} f_{0}(980)\to\pi^{0}\pi^{+}\pi^{-}$.

\begin{figure}[htbp]
       \centerline{\includegraphics[height=4cm,width=4 cm]
                                   {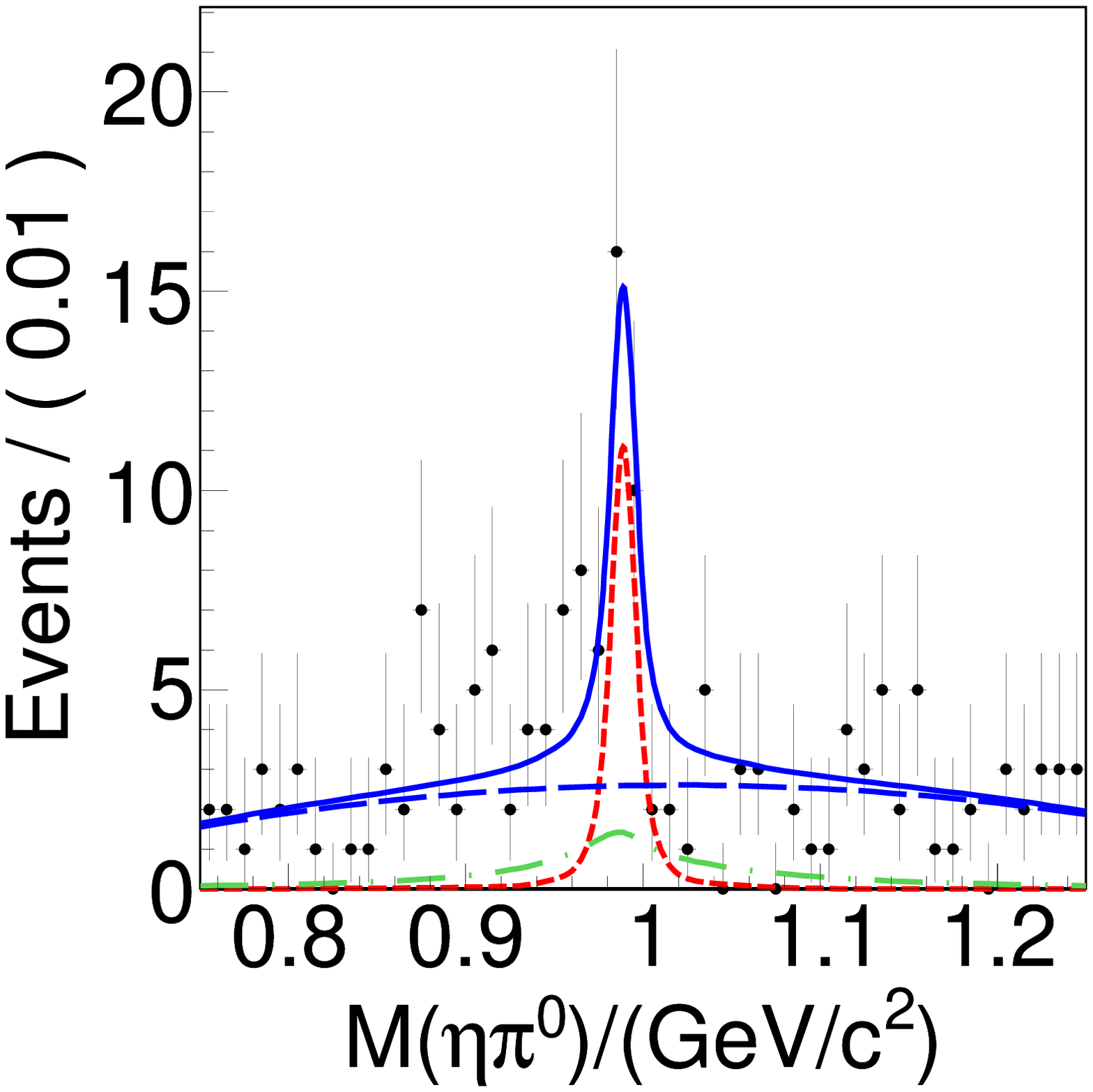}
                                   \includegraphics[height=4cm,width=4 cm]
                                   {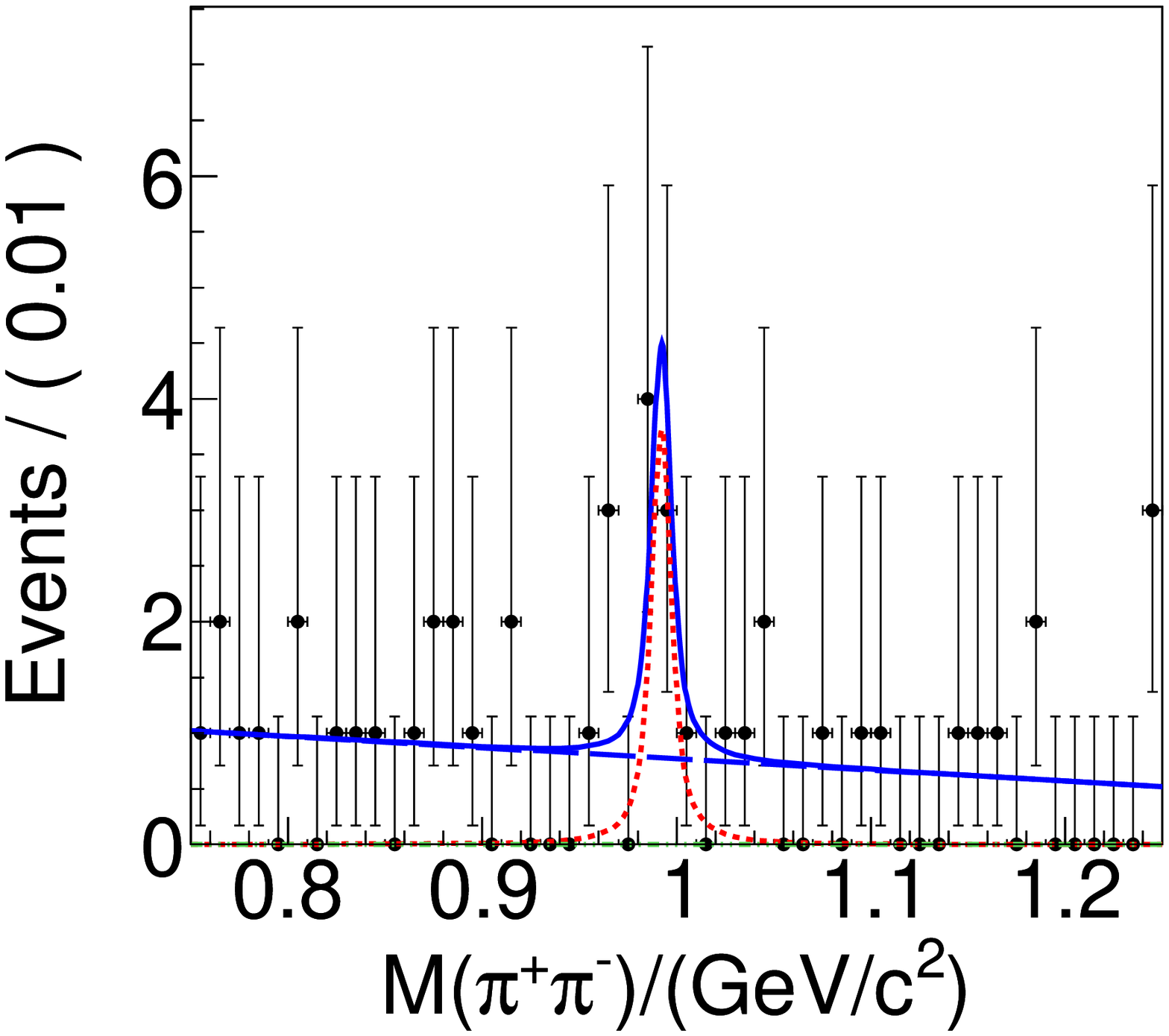}\put(-150,80){(a)}\put(-30,80){(b)}}
   \caption{(a)~Fitting result of the $\eta\pi^{0}$ mass spectrum recoiling against the $\phi$
   signal. (b)~Fitting result of the $\pi^{+}\pi^{-}$ mass spectrum in the $\chi_{c1}$ mass window.
   The dotted lines show the mixing signal. The dash-dotted lines
   indicate underlying $a^{0}_{0}(980)$ of $f_{0}(980)$ from other processes.
   The dashed lines denote the
   polynomial background.}
   \label{f0a0}
   \end{figure}

Figure~\ref{f0a0}~(a) shows the fitting results of the mass spectrum
of $\eta \pi^0$ recoiling against $\phi$ signal in $J/\psi\to\phi
f_{0}(980)\to\phi a^{0}_{0}(980)\to\phi\eta\pi^{0}$. The dots with
error bars are data. There is an evidence for a narrow peak over the
background in the expected resonance region. The dotted line is
signal. The dash-dotted line is $a_{0}(980)$ contribution from other
virtual photons and $K^{*}K$ loops. The dashed line is polynomial
background. The shape is constrained to the $\phi$ sideband. The fit
yields $N(f_{0}\to a^{0}_{0}) = 24.7\pm8.6$ (stat.) of $f_{0}(980)$
to $a_{0}(980)$ mixing signals with a significance of 3.3 $\sigma$.
The upper limit on the mixing branching ratio is determined to be
$Br(J/\psi\to\phi f_{0}(980)\to\phi
a^{0}_{0}(980)\to\phi\eta\pi^{0}) < 5.5\times 10^{-6}$ at the $90\%$
confidence level (C.L.).

Figure~\ref{f0a0}~(b) shows the fitting results of $\pi^{+}\pi^{-}$
invariant mass spectrum in $\chi_{c1}\to\pi^{0}
a^{0}_{0}(980)\to\pi^{0} f_{0}(980)\to\pi^0\pi^+\pi^-$. The fit to
the signal region is performed in a similar style as in the previous
$J/\psi\to\phi a_{0}(980)$ analysis. The fit yields $N(a^{0}_{0}\to
f_{0})=6.5\pm3.2$ (stat.) events for the mixing signal with a
significance of 2.0 $\sigma$. The upper limit on the mixing
branching ratio is determined to be
$Br(\psi^{\prime}\to\gamma\chi_{c1}\to\gamma\pi^{0}
a^{0}_{0}(980)\to\gamma\pi^{0}
f_{0}(980)\to\gamma\pi^{0}\pi^{+}\pi^{-}) < 5.5\times 10^{-7}$ at
the $90\%$ C.L.

The upper limit of the mixing intensity $\xi_{fa}$ for the
$f_{0}(980)\to a^{0}_{0}(980)$ transition at 90\% C.L. is calculated
to be:
\begin{eqnarray*}\label{eq_xifa}
\xi_{fa} = \frac {Br(J/\psi\to\phi f_{0}(980)\to\phi
a^{0}_{0}(980)\to\phi\eta\pi^{0})} {Br(J/\psi\to\phi
f_{0}(980)\to\phi\pi\pi)~\cite{phif0}}<1.1\% .
\end{eqnarray*}

The upper limit of the mixing intensity $\xi_{af}$ for the
$a^{0}_{0}(980)\to f_{0}(980)$ transition at 90\% C.L. is calculated
to be:
\begin{eqnarray*}\label{eq_xiaf}
\xi_{af} = \frac {Br(\chi_{c1}\to\pi^{0} a^{0}_{0}(980)\to\pi^{0}
f_{0}(980)\to\pi^{0}\pi^{+}\pi^{-})} {Br(\chi_{c1}\to\pi^{0}
a^{0}_{0}(980)\to\pi^{0}\pi^{0}\eta)~\cite{PDG}}
<0.9\%.\end{eqnarray*}

The measurements of $f_{0}(980)\to a^{0}_{0}(980)$ and $a^{0}_{0}(980)\to f_{0}(980)$ mixing transitions will be very helpful to probe the properties of
these two scalar states.

%Figure~\ref{intensity} shows the mixing intensities of
%$f_{0}(980)\to a^{0}_{0}(980)$ and $a^{0}_{0}(980)\to f_{0}(980)$
%and predictions with various theoretical and experimental values of
%the parameters. The dots are predictions for the mixing intensities
%with various theoretical and experimental values of the
%parameters~\cite{wujj2}. The shaded region is our measurement with
%error bars and the red lines are the upper limits. This result will
%be very useful for pinning down the resonance parameters of
%$a^{0}_{0}(980)$ and $f_{0}(980)$.

%   \begin{figure}[htbp]
%       \centerline{\includegraphics[height=4 cm,width=4 cm]
%                                   {intensity_2dim_01.eps}}
%   \caption{Mixing intensities of $f_{0}(980) \to a^{0}_{0}(980)$
%   and $a^{0}_{0}(980) \to f_{0}(980)$.The dots are predictions for the
%   mixing intensities with theoretical and experimental values of the
%   parameters~\cite{wujj2}. The shaded region is our measurement with
%   error bars. The red lines are our upper limits.}
%   \label{intensity}
%   \end{figure}

\section{Summary}
The charged $\kappa$ is observed at BESII in $J/\psi\to
K^{\pm}K_S\pi^\mp\pi^0$. Partial wave analysis on it gives the
consistent parameters with neutral $\kappa$. A new facility for
physics in the charm-$\tau$ region BEPC~II/BES~III has become
operational. With the world's largest samples of $J/\psi$ and
$\psi^\prime$ collected at BES~III, the direct measurement of $a_0$
and $f_0$ mixing is performed for the first time.

\end{document}